\documentclass[11pt]{article}
\usepackage{txfonts}
\usepackage{cospar}
\usepackage[sectionbib]{natbib}
\usepackage{url}
\usepackage{graphicx}

\catcode`\@=11
\newcommand{\gapprox}{\mathrel{\mathpalette\@versim>}}
\newcommand{\lapprox}{\mathrel{\mathpalette\@versim<}}
\newcommand{\@versim}[2]
  {\lower2.9truept\vbox{\baselineskip0pt\lineskip0.5truept
\ialign{$\m@th#1\hfil##\hfil$\crcr#2\crcr\sim\crcr}}}
\catcode`\@=12
\newcommand{\tbn}{\theta_{\rm Bn}}

\title{MICROPHYSICS OF SHOCK ACCELERATION FROM OBSERVATIONS OF
X-RAY SYNCHROTRON EMISSION FROM SUPERNOVA REMNANTS}

\author{S.P. Reynolds\address{Harvard-Smithsonian Center for Astrophysics,
60 Garden St., Cambridge, MA 02138\\ and \\
Physics Department, North Carolina State University,
Raleigh NC 27695-8202}}

\begin{document}

\maketitle

\begin{abstract}

Recent observations of non-thermal X-rays from supernova remnants have
been attributed to synchrotron radiation from the loss-steepened tail
of a non-thermal distribution of electrons accelerated at the remnant
blast wave.  In the test-particle limit of diffusive shock
acceleration, in which the energy in shock-accelerated particles is
unimportant, the slope of a shock-accelerated power-law is independent
of the diffusion coefficient $\kappa$, and on how $\kappa$ depends on
particle energy.  However, the maximum energy to which particles can
be accelerated depends on the rate of acceleration, and that does
depend on the energy-dependence of the diffusion coefficient.  If the
time to accelerate an electron from thermal energies to energy $E \gg
m_e c^2$ is $\tau(E)$, and if $\kappa \propto E^\beta$, then $\tau(E)
\propto E^\beta$ in parallel shocks, and $\tau \propto E^{2 - \beta}$
in perpendicular shocks. Most work on shock acceleration has made the
plausible assumption that $\kappa \propto r_g$ (where $r_g$ is the
particle gyroradius), so that $\beta = 1$ at relativistic energies,
implying a particular (wavelength-independent) spectrum of MHD
turbulence, where Kolmogorov or Kraichnan spectra might be more
physically plausible.  I derive the $\beta$-dependence of the maximum
electron energy resulting from limitations due to radiative
(synchrotron and inverse-Compton) losses and to finite remnant age (or
size).  I then exhibit calculations of synchrotron X-ray spectra, and
model images, for supernova remnants as a function of $\beta$ and
compare to earlier $\beta = 1$ results.  Spectra can be considerably
altered for $\beta < 1$, and images are dramatically different for
values of $\beta$ corresponding to Kolmogorov or Kraichnan spectra of
turbulence.  The predicted images are quite unlike observed remnants,
suggesting that the turbulence near SNRs is generated by the
high-energy particles themselves.

\end{abstract}

\section*{MICROPHYSICAL PARAMETERS OF STANDARD SHOCK-ACCELERATION
THEORY}

Diffusive (first-order Fermi) shock acceleration is presumed to
provide the relativistic-electron distributions demanded by radio and,
in a few cases, X-ray observations of supernova remnants (SNRs).  In
the test-particle limit in which the fast particles exert no influence
on the shock, the spectrum is independent of the value and
energy-dependence of the diffusion coefficient $\kappa$.  When the
energy in accelerated particles becomes non-negligible, however,
various nonlinear effects appear.  Most directly, the preshock gas is
pre-accelerated by a shock precursor whose extent depends on the
diffusion distance $r_D = \kappa/u_{shock}$ which does depend on
particle energies.  One observable effect is concave spectral
curvature if $\kappa(E)$ increases with $E$ (Ellison and Reynolds 1991;
Reynolds and Ellison 1992).  For both test-particle and nonlinear
acceleration, the acceleration {\sl rate} for relativistic electrons
depends on the diffusion coefficient.

Macrophysical parameters, those required for hydrodynamic and
thermal-shock modeling of SNRs, include the remnant age $t$, shock
velocity $u_{\rm sh}$ (ideally $u_{\rm sh}(t)$), preshock density
$n_0$, and compression ratio $r$.  (A piece of important thermal-shock
microphysics concerns non-Coulomb electron heating at shock, in the
absence of which $T_e \ll T_i$ initially).

For standard shock-acceleration theory, we need in addition the
upstream magnetic field $B_1$, the obliquity angle $\theta_{\rm Bn}$
between the shock normal and $B_1$, and the diffusion coefficient
upstream and downstream, $\kappa_1$ and $\kappa_2$ (in general
$\kappa(E, {\bf r}, t)$).  In the standard picture (e.g., Blandford and
Eichler 1987), in strong shocks ($M_A, M_s \gapprox 10$), once
particles have energies well above thermal, the primary scattering is
from resonant MHD waves (wavenumber $k \propto $ particle
gyrofrequency $\Omega$: $k_R = \Omega/c$). Quasi-linear theory allows
the inference of a diffusion coefficient. Assume a wave spectrum $I(k)
= Ak^{-n}$ erg cm$^{-3}$ (cm$^{-1})^{-1}$ (so the energy density at $k
\sim k I(k)$).  Then
\begin{equation}
\kappa_\parallel = {1 \over 3}\lambda_{\rm mfp} v 
= {c \over 3} \ r_g \ \left( {k_R I(k_R) \over {B^2 / 8\pi}}\right)^{-1}
\equiv {c \over 3} \ \eta(E) \ r_g
\end{equation}
where $\eta(E)$ is the (energy-dependent) ``gyrofactor,''
$\lambda_{\rm mfp} = \eta r_g$, and $\eta^{-1}$ is the fractional
magnetic energy in waves resonant with particles of that energy.  In
the so-called ``Bohm limit'', normally taken to give the minimum
physically possible diffusion coefficient, $\eta = 1.$

So $\eta \propto \left( k_R I(k_R) \right)^{-1} \propto k_R^{n-1}$.
Since $\Omega = eBc/E$ for extreme-relativistic particles, $\eta
\propto E^{1-n}$ and $\kappa_\parallel \propto \eta r_g \propto
E^{2-n} \equiv E^{\beta}$.  Then for Kolmogorov turbulence we have $n
= 5/3 \Rightarrow \beta = 1/3$, while for a Kraichnan turbulent
spectrum, $n = 3/2 \Rightarrow \beta = 1/2$.  Note that for constant
gyrofactor (as is often assumed), $\beta = 1 \Rightarrow n = 1$, {\it
i.e..} a ``white-noise'' spectrum with equal energy per decade.

This describes diffusion along magnetic-field lines.  Perpendicular
(cross-field) diffusion is generally assumed to result from a
cross-field displacement of one gyroradius for every scattering along
the field:
\begin{equation}
\kappa_\perp = {\kappa_\parallel \over 
            { 1 + \left( {\lambda_\parallel \over r_g} \right)^2 } }
= {\kappa_\parallel \over 
            { 1 + \eta^2 } }.
\end{equation}
Then 
\begin{equation}
\kappa = \kappa_\parallel \cos^2 \theta_{Bn} + 
               \kappa_\perp \sin^2 \theta_{Bn} 
  = \kappa_\parallel \left( \cos^2 \tbn + 
       {\sin^2 \tbn \over 1 + \eta^2 } \right).
\end{equation}
For quasi-perpendicular shocks ($\tbn$ near $90^\circ$) and large
$\eta$, $\kappa \ll \kappa_\parallel$ and acceleration can be much
faster than for parallel shocks (Jokipii 1987).  Even if $\eta = 1$
(Bohm limit), there is some obliquity dependence, mainly from field
compression where $\tbn \sim 90^\circ$.  But for large $\eta$, we have
{\sl faster} acceleration where the shock is perpendicular but {\sl
slower} where $\tbn$ drops well below $90^\circ$, giving larger
azimuthal brightness variations.  Large $\eta$ corresponds to weaker
turbulence, so that $\theta_{\rm Bn}$ is well-defined.  When $\eta$
is near 1, the obliquity-dependence is smaller anyway.

Radio observations of SNRs provide few constraints on the
microphysical parameters, because the power-law part of the spectrum
contains relatively little information: the diffusion-coefficent
dependence is only second order, through nonlinear effects (e.g., a
slight spectral hardening with energy; Ellison and Reynolds 1991). But
the maximum energies to which particles can be accelerated depend on
acceleration rates which strongly reflect the diffusion coefficient.
Now all known shell SNRs are fainter in X-rays (thermal or not) than
the extrapolation of their radio spectrum (14 Galactic remnants,
Reynolds and Keohane 1999; 11 LMC remnants, Hendrick and Reynolds 2001).
So any non-thermal emission results from the cutting-off tail of the
electron distribution, giving direct information on the maximum
energies.  Such synchrotron X-ray emission has now been identified in
several Galactic remnants, inferred from lineless spectra that cannot
be produced by thermal processes (SN 1006: Koyama et al.~1995;
G347.5-0.3, Slane et al.~1999; and several others).

These remnants can then be used to study various microphysical
quantities.  The magnitude of the diffusion coefficient is constrained
by the maximum electron energies observed, and its upstream value in
particular is related to possible ``halo'' emission from upstream
electrons diffusing ahead of the shock.  The energy-dependence of the
diffusion coefficient can be constrained by studying remnant
morphology and detailed spectral shape, compared with model
calculations such as those to be summarized below.  The obliquity
dependence of acceleration affects the azimuthal variation of X-rays
around the shell.  The overall compression ratio can be constrained by
the halo visibility (and thermal modeling).  Finally, observed maximum
energies also constrain the shock speed and other hydrodynamic
parameters.

\section*{SYNCHROTRON X-RAY OBSERVATIONS OF REMNANTS}

SN 1006, the prototype object for synchrotron X-rays, has a spectrum
that is well fit by a model in which acceleration is limited by
electron escape above some $E_{\rm max}$ (Reynolds 1996).  The fitted
characteristic rolloff frequency of $7 \times 10^{17}$ Hz (where the
spectrum has dropped a factor $\sim 6$ below the lower-frequency
extrapolation; Long et al.~2002), with the known mean magnetic field
(from inverse-Compton TeV emission, $\langle B \rangle \sim 10 \
\mu$gauss; Tanimori et al.~1998; Dyer et al.~2001) implies $E_{\rm
max} = 53 $ TeV.  Furthermore, the known shock velocity implies that
the observed electron spectral cutoff energy is too low to result from
synchrotron losses in this magnetic field ($t_{1/2}(E_{\rm max})
\gapprox 1200$ yr) or finite age ($t_{\rm accel} \lapprox 800$ yr).
This leaves only a change in diffusive properties for $\lambda_{\rm
MHD} \gapprox 2 \times 10^{17}$ cm to explain the observed rolloff.

The detailed cylindrically symmetric model of Reynolds (1996; 1998)
fits the spectrum well (Dyer et al.~2001), but predicts an external
halo with brightness 5 -- 10\% of postshock peak just ahead of the
shock, and an extent of many arcmin, the diffusion length of the
high-energy electrons.  However, no halo is seen in the deep,
low-background {\it Chandra} image: any preshock emission $\lapprox$
1\% of the postshock maximum on the bright rims.  Plane-shock models
seen edge on predict a lower brightness, $\sim $3\%, but still in
conflict with the observations.  While efficient acceleration of
particles can increase the overall compression ratio (e.g., Berezhko,
Elshin, et al., 1996), this occurs over the long diffusion
lengthscale, while the immediate density jump at the small-scale
thermal subshock actually decreases, making the problem worse.  This
problem is discussed at length in Long et al.~(2002) where it is
suggested that the post-shock magnetic field may actually increase by
more than due to simple fluid compression, perhaps from some kind of
cosmic-ray-driven instability (Lucek and Bell 2000).

In general, the halo can be made unobservable by making it thin rather
than faint. For this option, the diffusion length $\kappa_1/u_{\rm
sh}$ must be less than the width of the sharp edge, $\lapprox
10^{\prime\prime}$ for the {\it Chandra} observations.  This would
require an unreasonably high upstream magnetic field, at least 20 
$\mu$G, especially given SN 1006's height of 570 pc above the
Galactic plane.

\begin{figure}
\includegraphics[width=88mm]{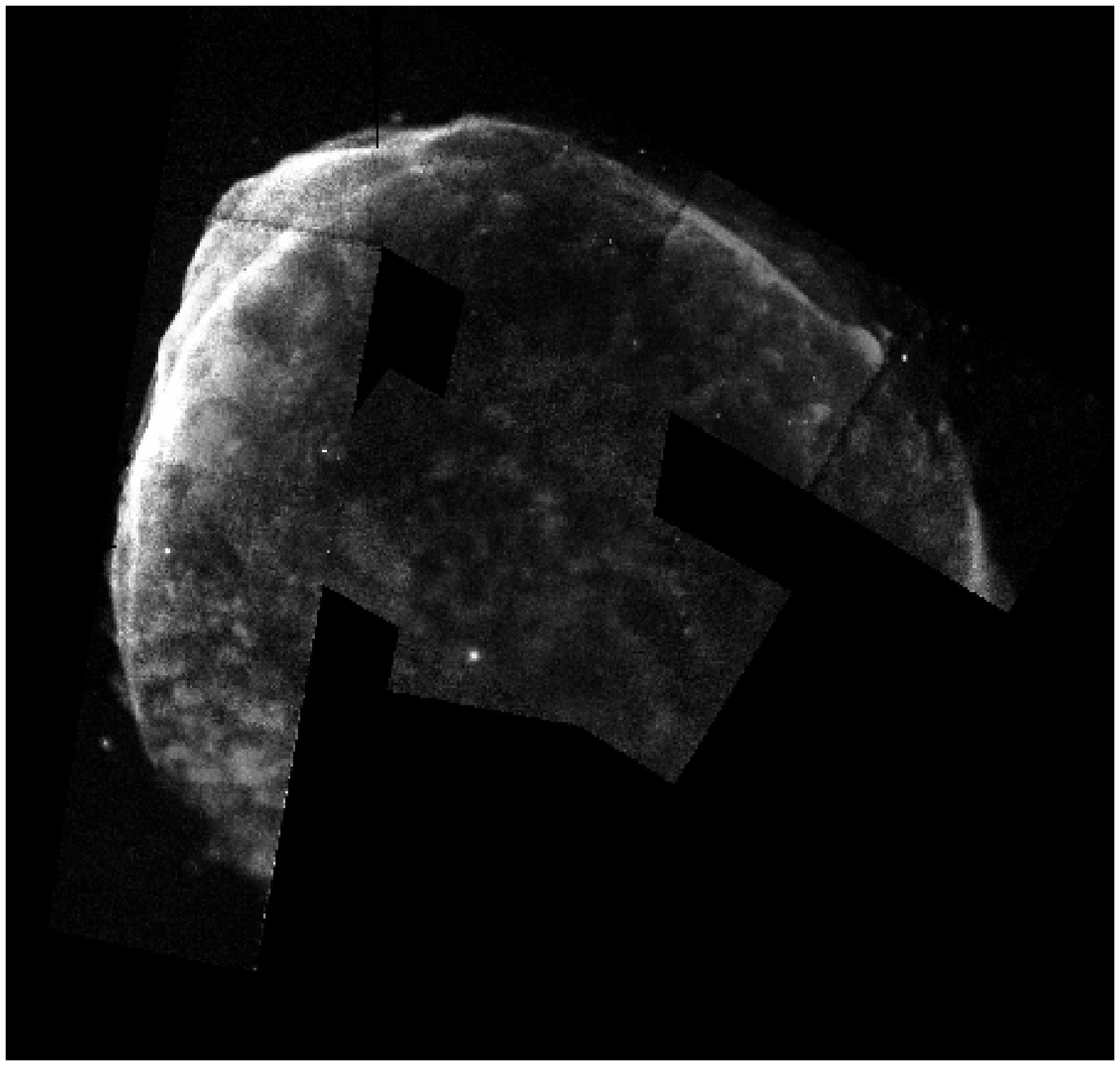}\quad
\includegraphics[width=86mm]{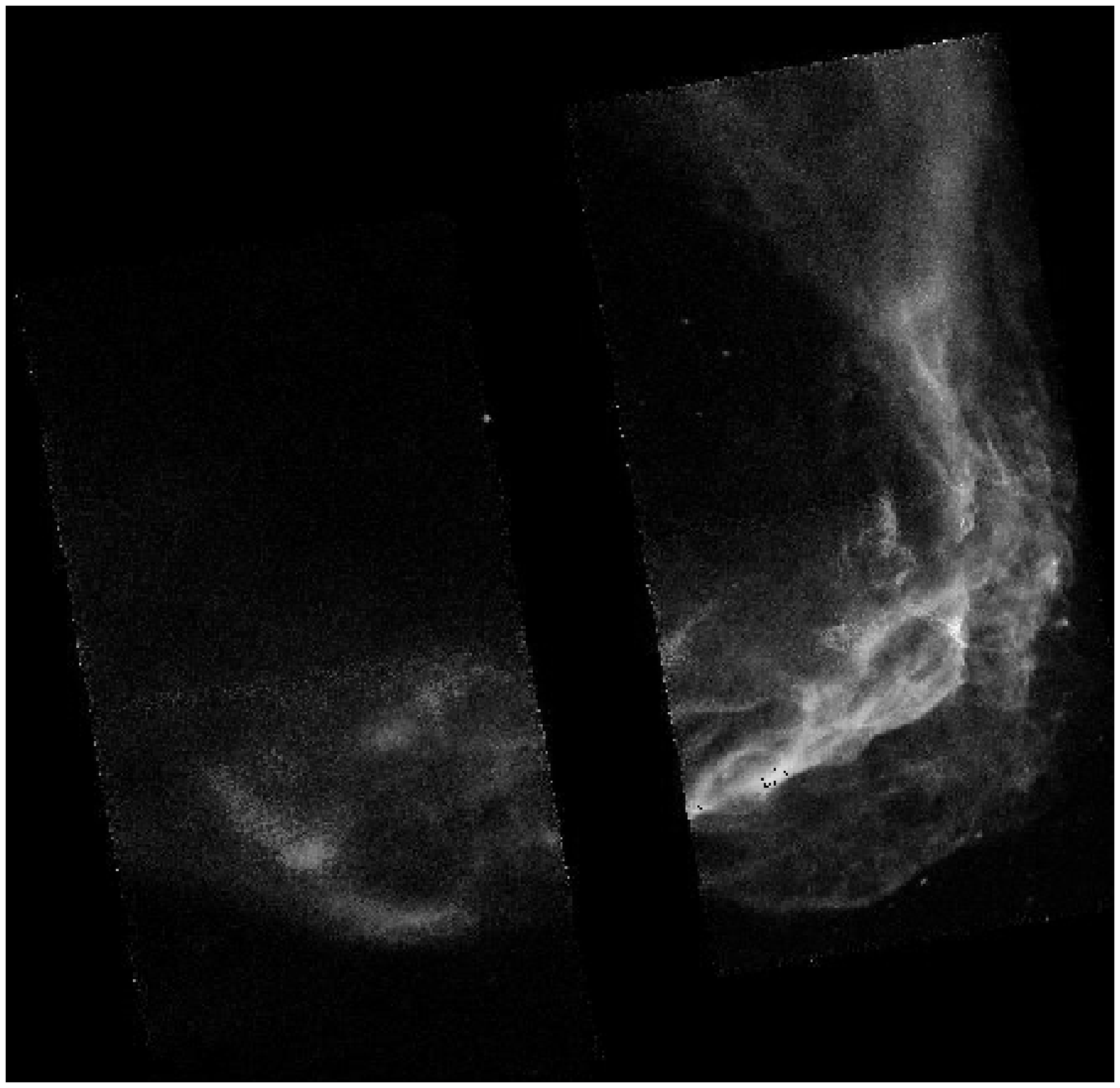}
\caption{Left: SN 1006 {\it Chandra} image (adapted from Long et 
al.~2002).  Right: RCW 86 (SW corner), {\it Chandra} image (adapted 
from Rho et al.~2002).} 
\end{figure}

In the larger, older remnant RCW 86, weak X-ray lines in SW corner
imply anomalous abundances if the continuum is thermal (Vink et
al.~1997), while energetics problems (Rho et al.~2002) afflict an
explanation of the continuum as non-thermal bremsstrahlung (Vink et
al.~2002).  However, if the harder continuum is dominated by
synchrotron emission, reasonable abundances can be accommodated.
(Borkowski et al.~2001; Rho et al.~2002).  {\it Chandra} data imply
rolloff frequencies $\nu_c \sim (7 - 10) \times 10^{16}$ Hz,
consistent with loss-limited acceleration (shock velocities are known
from H$\alpha$ emission to be 600 -- 900 km s$^{-1}$; Ghavamian et
al.~2002).  For radiative-loss-limited acceleration,
\begin{equation}
\nu_c = 5 \times 10^{16} \ \eta \ {B_2 \over 4B_1} \  
\left({u_{\rm sh} \over 1000 \ {\rm km s}^{-1}}\right)^2 \ {\rm Hz}
\end{equation}
so we need $\eta > 1$ and/or \ $r > 4$ again, even for a perpendicular
shock.

\section*{CONSTRAINTS ON ENERGY DEPENDENCE OF THE DIFFUSION COEFFICIENT}

I have generalized the modeling code of Reynolds (1998) to include
arbitrary energy-dependence of the diffusion coefficient.  The models
assume a spherical blast wave encountering uniform upstream material
containing a uniform magnetic field whose sky-plane projection is
vertical and which makes an angle $\phi$ with respect to the sky
plane.  The remnant dynamics are Sedov; each post-shock fluid element
is endowed with an electron distribution $N(E) = K E^{-s}
\exp(-E/E_{\rm max})$, where the maximum energy is in general a
function of shock speed, remnant age, diffusion coefficient, and shock
obliquity angle $\tbn$.  This distribution is then evolved downstream
subject to radiative and adiabatic losses; electrons are assumed to
have short enough post-shock mean free paths that they stay confined
to their original fluid element and share its expansion.  Details are
given in Reynolds (1998).

Here we generalize by allowing an energy dependence of the diffusion
coefficent, $\kappa \propto E^\beta$.  For the relatively steep
spectra of plasma waves predicted in simple homogeneous turbulence
models, such as Kolmogorov turbulence, $\beta < 1$, so that
$\lambda_\parallel$ rises more slowly with energy than $r_g$.  At some
high energy $E_h,$ then, we will have $\lambda_\parallel = r_g$.
Above this energy, we must have an effective $n = 1$ (constant
``gyrofactor.'')  So we confine our attention to $E < E_h$ and write
\begin{equation}
\kappa = \kappa_B 
\left( {E \over E_h} \right)^\beta \qquad \qquad 
{\lambda_\parallel \over r_g} = \left( {E \over E_h} \right)^{\beta - 1}
\qquad \qquad E < E_h
\end{equation}
where $\kappa_B = r_g(E_h) c/3 = E_h c/3eB$, the Bohm-limit value at
$E = E_h$. 

Then for a parallel shock, $\theta_{Bn} = 0^\circ$, the acceleration
time is given by (generalizing from Reynolds 1998)
\begin{equation}
\tau_{acc} = {{r (r+1)} \over {\beta(r-1)}} 
   {c \over e} {E_h \over u_1^2 B_1} \left( {E \over E_h} \right)^\beta
\end{equation}
whereas for a 
\noindent
perpendicular shock, $\theta_{Bn} = 90^\circ$, we have
\begin{equation}
\tau_{acc} = {{2 r } \over {(2 - \beta)(r-1)}}\  
   {c \over e} \ 
   {E_h \over u_1^2 B_1} \left( {E \over E_h} \right)^{2 - \beta}
\end{equation}
with $r$ the shock compression ratio.  Note that
$\tau(90^\circ)/\tau(0^\circ) \propto (E / E_h)^{(2-2\beta)}$ so that
if $\beta < 1,$ acceleration to energies $E \ll E_h$ is much faster at
perpendicular shocks than parallel.

The result of these effects for supernova-remnant morphology is
dramatic (see Figure 2).  For a spherical remnant expanding into a
uniform magnetic-field $B_1$, around the ``equator'' where
$\theta_{Bn} \sim 90^\circ$ the electron spectrum will continue to
much higher energies than near the ``poles'' ($\theta_{Bn} \sim
0^\circ$) if $\beta < 1$.  However, the spectrum will not differ much
from the $\beta = 1$ case because the emission is dominated by the
bright equatorial ``belt''.

\begin{figure}
\centerline{\includegraphics[width=110mm]{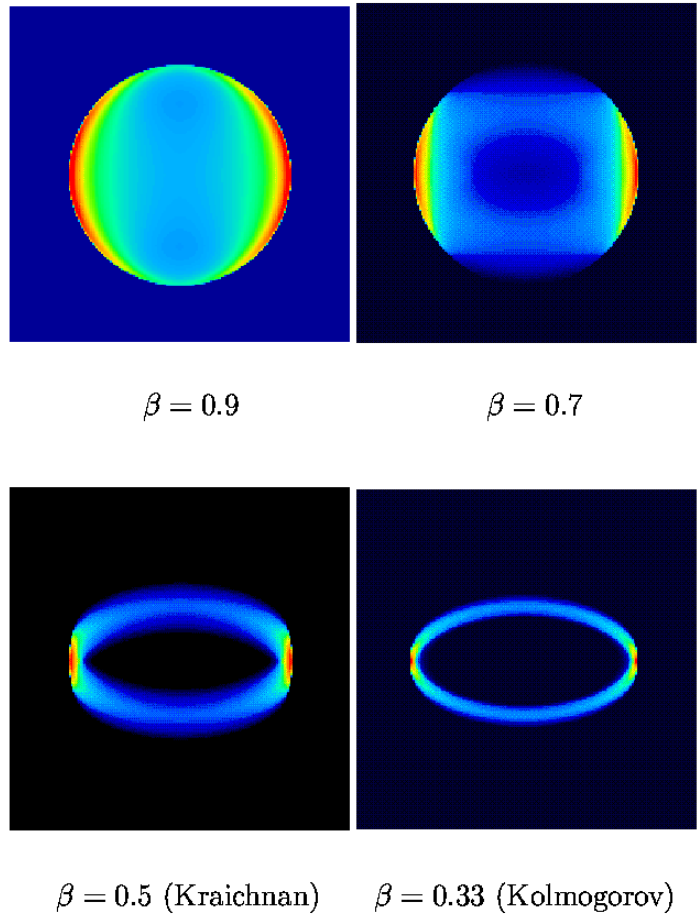}}
\caption{Four simulations for different values of $\beta$. The magnetic
field is assumed to have a vertical projection on the sky plane, and
to make an angle $\phi = 30^\circ$ with that plane.}
\end{figure}

\section*{CONCLUSIONS AND FUTURE WORK}

Because it samples electron energies where the spectrum is cutting
off, X-ray synchrotron emission from SNRs can provide significant
information on the microphysics of shock acceleration (mainly on the
diffusion coefficient), to a much greater extent than radio
synchrotron emission.  In the remnant of SN 1006, the shock is almost
certainly perpendicular at the bright non-thermal rims.  Simple
predictions for the spectral break due to radiative losses give too
high a break, as do predictions based on the finite remnant age (or
size).  For the spectrum to break where it is observed, the diffusive
properties of the external medium probably have to change on scales
above $\sim 0.1$ pc, so that electrons escape much more freely when
their gyroradii are comparable to this scale.  The absence of an
exterior synchrotron ``halo'' probably requires a cross-shock jump in
magnetic-field strength considerably greater than given by the fluid
compression ratio there, suggesting the possibility of some kind of
immediate post-shock turbulent or plasma amplification of magnetic
field.  In the considerably older remnant RCW 86, synchrotron emission
is important, though not dominant; careful modeling of both thermal
and non-thermal emission is required to extract information.  The
rather high observed break frequencies require rapid acceleration:
perpendicular shocks with $\eta > 1$ and/or $r > 4$.

The modeling of shock acceleration for diffusion in steep wave spectra
in the context of a spherical SNR shock has shown that
quasi-perpendicular favoritism, already present in constant-gyrofactor
models, is much more extreme if $\beta < 1$ (wave spectra steeper than
$k I_k \propto k^{-1}$).  While predicted spectra aren't much
different from the $\beta = 1$ (constant-gyrofactor) case, predicted
images are strikingly unlike real remnants for $\beta < 1$.  A likely
explanation is that the MHD waves causing diffusion are
self-generated, with $k^{-1}$ spectrum.  In general, however, it is
clear that we need better thermal and non-thermal models for X-ray
emission in SNRs.  Such improved models ought to allow the extraction
of a great deal of important and useful information about the
microphysics of shock acceleration.

\vskip 0.3 truein

\noindent
e-mail: steve\_reynolds@ncsu.edu

\vfill

\end{document}